\newif\iftechreport
\techreportfalse 
\iftechreport
	\documentclass[12pt,technote,draftclsnofoot,onecolumn]{elsarticle}
\else
\documentclass[review,natbib]{elsarticle}
\fi

\usepackage{float}
\usepackage{amssymb}
\usepackage{amsmath}
\usepackage{mathtools}
\usepackage[caption=false,font=footnotesize]{subfig}
\usepackage{color}
\usepackage[none]{hyphenat}
\usepackage{booktabs}
\usepackage{multirow}
\usepackage{url}

\usepackage{natbib}
\biboptions{authoryear}
\usepackage{lineno}
\modulolinenumbers[5]

\usepackage{breqn,ulem}
\usepackage{tikz}
	\usetikzlibrary{shapes,arrows}
\usepackage{pbox}
\usepackage{blindtext}
\usepackage{array}
\usepackage[ruled,vlined]{algorithm2e}
\DeclareMathOperator*{\argmin}{arg\,min}
\DeclareMathOperator{\prox}{\mathrm{prox}}
\usepackage{url}
\usepackage{hyperref}
\journal{a journal}
\begin{document}
\begin{frontmatter}
\title{Compartment Model-based Nonlinear Unmixing for Kinetic Analysis of Dynamic PET Images}

\author[n7]{Yanna Cruz Cavalcanti\corref{cor1}}
\ead{yanna.cavalcanti@enseeiht.fr}
\author[isae]{Thomas Oberlin}
\ead{thomas.oberlin@isae-supaero.fr}
\author[n7]{Vinicius Ferraris}
\ead{vinicius.ferraris@enseeiht.fr}
\author[n7,iuf]{Nicolas Dobigeon}
\ead{nicolas.dobigeon@enseeiht.fr}
\author[umrs]{Maria Ribeiro}
\ead{maria.ribeiro@univ-tours.fr}
\author[umrs]{Clovis Tauber}
\ead{clovis.tauber@univ-tours.fr}
\address[n7]{University of Toulouse, IRIT/INP-ENSEEIHT, 31071 Toulouse Cedex 7, France}
\address[isae]{ISAE-SUPAERO, University of Toulouse, France}
\address[iuf]{Institut Universitaire de France, France}
\address[umrs]{UMRS Inserm U930 - Universit\'e de Tours, 37032 Tours, France}
\cortext[cor1]{Corresponding author}
\tnotetext[fund]{Part of this work has been funded by the ANR-3IA Artificial and Natural Intelligence Toulouse Institute (ANITI).}

\begin{abstract}
When no arterial input function is available, quantification of dynamic PET images requires a previous step devoted to the extraction of a reference time-activity curve (TAC). Factor analysis is often applied for this purpose. This paper introduces a novel approach that conducts a new kind of nonlinear factor analysis relying on a compartment model, and computes the kinetic parameters of specific binding tissues jointly. To this end, it capitalizes on data-driven parametric imaging methods to provide a physical description of the underlying PET data, directly relating the specific binding with the kinetics of the non-specific binding in the corresponding tissues. This characterization is introduced into the factor analysis formulation to yield a novel nonlinear unmixing model designed for PET image analysis. This model also explicitly introduces global kinetic parameters that allow for a direct estimation of the binding potential with respect to the free fractions in each non-specific binding tissue. The performance of the method is evaluated on synthetic and real data to demonstrate its potential interest.
\end{abstract}
\end{frontmatter}

\begin{keyword}
dynamic PET, factor analysis, nonlinear unmixing, binding potentials.
\end{keyword}
    \vspace{-0.5cm}

\section{Introduction}
Dynamic positron emission tomography (PET) is an imaging technique delivering relevant information on {physiological} dysfunctions preceding appearance of morphological abnormalities such as cancer and dementia. To provide interpretable results, PET images have to pass through a process called quantification \citep{BUVAT2007}. It consists in exploring the variations of the concentration of radiotracers over time, characterized by time-activity curves (TACs), in order to estimate the kinetic parameters that describe the studied process. {Most of the quantification techniques} {that are used in practice} require an estimation of reference TACs representing tissues \citep{blomqvist1989dynamic,cunningham1991compartmental}. In this context, many methods were developed to perform a non-invasive extraction of the global kinetics of a tracer, in particular factor analysis \citep{Barber1980,Cavailloles1984}.

However, standard factor analysis techniques {rely} on the  assumption that the elementary response of each tissue to tracer distribution is spatially homogeneous. To overcome this limitation,  \cite{unmixingCavalcanti2018} introduced a factor analysis model that handled fluctuations in specific binding (SB) kinetics with a spatially indexed variability. The proposed model tried to extract a factor for the blood input function, a factor for each non-specific binding (nSB) tissue of the region under study and assigned a spatially varying factor for high-uptake tissues.

Furthermore, the kinetics of SB tissues are often related to that of nSB tissues, as considered by the reference tissue input models from the quantification literature \citep{lammertsma1996simplified}. Indeed, when {targets of the radiotracer} are present, the kinetics of a tissue that is non-specific under healthy circumstances will be nonlinearly modified in the presence of the labelled molecule, though it is still the same tissue or organ.
Therefore, this work proposes to study SB as an instance of nSB kinetics. The main idea is to perform factor analysis on nSB tissues and blood while allowing for nonlinearities on each nSB tissue that will describe the {SB part present in these regions}.  

Ideally, these nonlinearities should directly provide an interpretable result in terms of quantification, representing different levels of binding. To do so, one resorts to the parametric pharmacokinetic models discussed by \cite{gunn2001positron}. They are shown to be very useful for providing physiologically meaningful {analysis of PET data}. Conventional methods used for kinetic parameter estimation often define a region-of-interest (ROI) and then perform estimation based on the ROI average \citep{lammertsma1996simplified,Berti2010}. These approaches neglect any spatial variations in the tracer kinetics within the ROI, e.g., due to partial volume effects and tissue heterogeneity. To avoid this homogeneous ROI assumption, some studies performed a voxel-by-voxel estimation of kinetic parameters and, to overcome poor signal-to-noise ratio (SNR), applied additional penalizations to stabilize the estimation \citep{Kamasak2005,Huang1998}. For instance, \cite{Gunn2002} defined basis functions to model the kinetics of the tracer. The basis functions are pre-computed and an undetermined system of equations is {solved} to fit to the data with a technique named DEPICT. Since this is an ill-posed problem, an additional sparsity penalization is imposed to the basis coefficients. This sparsity assumption is motivated by the fact that data is expected to be described by a few compartments. \cite{Peng2008} investigated the use of sparse Bayesian learning for parametric estimation{, further allowing the coefficients to be nonnegative, in agreement with reference tissue models.}

Nevertheless, these approaches still assume that there is only one kinetic process occurring per voxel whereas, due to the low spatial resolution, the partial volume effect and the biological heterogeneity, the resulting signal is often a mixture of multiple kinetic processes. This is also the rationale behind several factor analysis models proposed in the literature \citep{dipaola1982,Sitek2000}. To overcome this limitation, \cite{Lin2014} proposed a two-stage algorithm that benefits from prior information provided by parametric imaging models on the physics and physiology of metabolism while introducing partial volume with a linear combination of the different kinetics. The first step consisted in a dictionary-based estimation of the nonlinear kinetics of each considered tissue and the second step computed the tissue fractions and the linear terms of the tissue kinetic models. This model considers that each image voxel is described by a linear mixing of $K$ classes, including blood, and assumes that the blood input function $\mathbf{m}_K$ is known. Each tissue factor TAC $\mathbf{m}_k(\boldsymbol{\kappa})$, for $k=1,\cdots,K-1$, is described by a three-tissue compartment model \citep{Huang1980} with kinetic parameters $\boldsymbol{\kappa}$. The final formulation for each voxel can be written as
\begin{equation}
\mathbf{x}_n = \sum_{k=1}^{K-1} a_{kn}\mathbf{m}_{k}(\boldsymbol{\kappa})+a_{Kn}\mathbf{m}_K.
\label{eq5:lin}
\end{equation}
A similar approach was also proposed by \cite{Chen2011}.  {Based on the same idea, \cite{Klein2010} tries to describe each factor TAC with an input function-based kinetic model and to jointly estimate this input function as well as the model parameters for each factor.}

However, many experimental results indicate that the use of commonly accepted multi-compartment models often leads to considerably biased and high-variance estimates of the pharmacokinetics parameters, due to the high number of parameters to be estimated \citep{padhani2001dynamic,padhani2003mri,buckley2002uncertainty}. Moreover, they often oversimplify the kinetics of several tracers, especially in case of tissue heterogeneity \citep{delorenzo2009modeling}. As an attempt at providing a more accurate description of the kinetics of the tracer while benefiting from the physiological description of parametric imaging, the approach proposed in this paper relies on a parametric nonlinear factor analysis. Differently from the approach followed by \cite{Lin2014}, factor TACs from nSB tissues will be directly estimated in the model and, based on the data-driven reference input model by \cite{Gunn2002}, will be used as reference tissue TACs for the recovery of the kinetic parameters from SB. 
{The idea of linking factor analysis to compartmental modeling has already been investigated by some works from the PET literature. In particular, \cite{nijran1985} proposed to constrain the space of possible solutions of factor analysis with the space of theoretical solutions given by compartmental models. \cite{Szabo1993} used factor analysis to differentiate SB and nSB TACs and determine the number of compartments needed to model the kinetics of [11C] pyrilamine in the brain. \cite{Elfakhri2005,ElFakhri2009} used a previous factor analysis step to extract the input functions that were used to compute the TACs in each voxel with a two-compartment model.
The work proposed in this article goes one step further by jointly conducting factor and kinetic analysis.}
This paper is organized as follows. Section \ref{sec:background} provides the theory that underlies this work. The proposed
{nonlinear model and the corresponding unmixing framework are introduced in Section \ref{sec5:prob_stat}}.
Results from simulations conducted on synthetic and real image are reported in Section \ref{sec5:synt_results} and  Section \ref{sec5:real_results}, respectively. Section \ref{sec5:conclu} concludes the {paper}.\vspace{-0.15cm}

\section{Background}
\label{sec:background}
\subsection{Nonlinear mixing models}\label{subsec:nonlinearmodels}
Let $\mathbf{x}_n \in \mathbb{R}^L$ be the $n$th voxel TAC neglecting any measurement noise, where $L$ is the number of time-frames. Linear mixing models (LMM), which are the basis of factor analysis, assume each TAC to be a linear combination of $K$ elementary factors $\mathbf{m}_k \in \mathbb{R}^L$ and their respective mixing coefficients $a_{kn}$ \citep{Bioucas-Dias2012,Dobigeon2009}. More explicitly, this model is mathematically expressed as
\vspace{-0.15cm}\begin{equation}
\mathbf{x}_n = \sum_{k=1}^K{a_{kn}\mathbf{m}_k} = \mathbf{M}\mathbf{a}_n\vspace{-0.15cm}
\label{eq:FA_pixel}
\end{equation}
with $\mathbf{M}= \left[\mathbf{m}_1,\ldots,\mathbf{m}_K\right]$ and $\mathbf{a}_n=\left[a_{1n},\ldots,a_{Kn}\right]$. Moreover, {non-negativity} constraints are assumed for the factors $\mathbf{m}_k=\left[{m}_{1,k},\ldots,{m}_{l,K}  \right]^T$, i.e.,
\vspace{-0.15cm}\begin{equation}
{m}_{lk} \geq 0 \quad 
\label{eq:constrM}\vspace{-0.15cm}
\end{equation}
while factor proportions are constrained to be {non-negative} and sum to one
\vspace{-0.15cm}\begin{equation}
{a}_{k,n} \geq 0 \quad \text{and} \quad 
\sum_{k=1}^K {a}_{kn}  = 1.\vspace{-0.15cm}
\label{eq:constrA}
\end{equation}
When nonlinearities are present, this model is no longer sufficient to describe the {behavior of the TACs}. 
To {overcome this issue}, several nonlinear mixing models and their corresponding unmixing algorithms were proposed, raising a fertile branch of research in hyperspectral imaging {for Earth observation} \citep{heylen2014review,dobigeon2014nonlinear,Dobigeon_IEEE_JSTARS_2014}.
A large family of nonlinear models  can be described as \citep{Altmann_IEEE_WHISPERS_2011,Meganem2014,Eches2014}
\vspace{-0.15cm}\begin{equation}
\mathbf{x}_n=\mathbf{M}\mathbf{a}_n+\boldsymbol{\mu} (\mathbf{M},\mathbf{a}_n,\mathbf{b}_n),\vspace{-0.15cm}
\label{eq:nonlin_unmix}
\end{equation}
where, in addition to the linear contribution in \eqref{eq:FA_pixel}, the observed pixel is also composed of an additive nonlinear term $\boldsymbol{\mu}(\cdot)$ that may depend on the factors matrix $\mathbf{M}$, the factor proportion coefficients $\mathbf{a}_n$ and {new} additional nonlinearities coefficients $\mathbf{b}_n$.
The rationale behind the relation between the nonlinear coefficients and the amount of linear contributions $a_{kn}$ comes from the fact that a pixel containing more of a given factor is more subjected to nonlinear interactions. In other words, if a material is not present in one pixel, it cannot interact with other materials \citep{fan2009comparative}. \vspace{-0.2cm}

\subsection{Reference tissue input models}
Tracer kinetic modelling techniques are used to estimate biologically interpretable parameters by describing the TACs in a ROI with mathematical models. A wide range of techniques model the PET signals based on compartmental analysis of the tracer \citep{gunn2001positron}. These approaches may be divided into two major groups: model-driven methods and data-driven methods \citep{Gunn2002}. Model-driven methods are based on a previously chosen compartmental model, whereas data-driven techniques do not need any \textit{a priori} decisions about the most appropriate model structure.

In this work, we will be specially interested in the general model based on reference tissues developed by \cite{Gunn2002}. It relies on a basis function framework and it does not require any knowledge on the compartment model. Each target voxel TAC is described as
\vspace{-0.15cm}\begin{equation}
\mathbf{x}_n(\mathbf{b}_n,\boldsymbol{\alpha}) = \left((1+b_{0n})\delta(\mathbf{t})+\sum_{i = 1}^{V}b_{in}  e^{-\alpha_{i}\mathbf{t}}\right) \ast \mathbf{m}_{\mathrm{R}},\vspace{-0.15cm}
\label{eq5:gunnFRCM}
\end{equation}
where $V$ is the total number of tissue compartments in both the target and reference tissues, $\mathbf{t} = [t_1,\cdots,t_L]^T$ are the times of acquisition which are assumed to be known, $\ast$ stands for temporal convolution, $\delta(\mathbf{t})$ is the Dirac impulse function and $e^{\boldsymbol{\theta}}$ is a point-wise exponentiation. {The tracer delivery ratio between the $n$th targeted voxel and the reference region is given by ${R}_{1{n}}= 1+b_{0n}$.}
{As before, $\mathbf{x}_n$ denotes the TAC in the $n$th voxel, while $\mathbf{m}_{\mathrm{R}}$ is the reference tissue TAC of the studied ROI, and the new variables $\mathbf{b}_n=[b_{1n},\ldots,b_{Vn}]$ and $\boldsymbol{\alpha}=[\alpha_1,\ldots,\alpha_V]$ describe the kinetics of the tracer.}
This formulation neglects the blood volume in both target and reference tissues. 
For models with reversible target tissue kinetics ($\alpha_{i}>0$), \cite{gunn2001positron} derived a direct relation {between} the computed kinetic parameters {and} the binding potential (BP) with respect to the free fractions of the radioligand in tissue ($f_{\mathrm{T}}$), denoted $\mathrm{BP}.f_{\mathrm{T}}$ and given by
\vspace{-0.15cm}\begin{equation}
\mathrm{BP}.f_{\mathrm{T}} = b_{0n}+\sum_{i = 1}^{V}\frac{b_{in}}{\alpha_{i}}.\vspace{-0.15cm}
\label{eq1:binding_potential}
\end{equation}
\vspace{-0.2cm}

\section{Proposed framework}
\label{sec5:prob_stat}

\subsection{{Nonlinear PNMM model}}
The proposed framework combines the model in \eqref{eq5:gunnFRCM} with the generalization of the linear mixing \eqref{eq5:lin} by associating each reference TAC to a nSB tissue TAC, except for the blood factor TAC $\mathbf{m}_K$. This yields the  following so-called parametric nonlinear mixing model (PNMM)
\vspace{-0.15cm}
\begin{align}
\mathbf{x}_n = \sum_{k=1}^{K-1} a_{kn}\left(\mathbf{m}_{k}+\sum_{i = 1}^{V}b_{kin}\mathbf{m}_{k}\ast  e^{-\alpha_{ki}\mathbf{t}}+b_{k0n}\mathbf{m}_{k}\right) 
+a_{Kn}\mathbf{m}_K\vspace{-0.15cm}
\end{align}
where $b_{kin}$ receives an additional index $k$ since this coefficient is {now specific to each reference tissue TAC}. This model is expected to be more flexible than the previous ones, since it accounts for possible partial volume effects induced by mixing between tissues and blood, while benefiting from the physical considerations of parametric imaging. It also directly estimates the global kinetics of one tissue, thus not being completely dependent of kinetic parameters. This can offer a more precise quantification as it automatically analyses different nSB tissues separately, {accounting implicitly for possible differences in perfusion of such tissues}. It may also allow the tissue affected by SB to be identified, through the computation of the BP within each nSB tissue. 

Adopting matrix notations consistent with those introduced in Section \ref{subsec:nonlinearmodels}, the noise-free image $\mathbf{X} \in \mathbb{R}^{L \times N}$ writes
\vspace{-0.15cm}\begin{equation}
\mathbf{X} = \mathbf{M}\mathbf{A}+\sum_{i=0}^V\mathbf{Q}_i(\mathbf{\tilde{A}}\circ \mathbf{B}_i)\vspace{-0.15cm}
\label{eq5:pnmm}
\end{equation}
where  $\mathbf{M}$ is a $L \times K$ matrix containing the factor TACs, $\mathbf{A}$ is a $K \times N$ matrix composed of the factor proportion vectors $\mathbf{a}_n=\left[a_{1n},\ldots,a_{Kn}\right]^T$ and $\circ$ is the Hadamard point-wise product.
{The matrix $\mathbf{\tilde{A}}$ denotes the factor proportion matrix while omitting the blood, i.e., $\mathbf{\tilde{A}} = [\mathbf{a}_1,\cdots,\mathbf{a}_{K-1}]$. The kinetic parameters are encoded in the matrices $\mathbf{Q}_i$, with $i\in\{0,\ldots,V\}$, depending on $\mathbf{M}$ and $\boldsymbol{\alpha}$,}
\vspace{-0.1cm}\begin{equation}\label{eq:Qi_1}
\begin{split}
\mathbf{Q}_i=\begin{bmatrix}
\mathbf{m}_1\ast e^{-\alpha_{1i}\mathbf{t}}&\cdots&\mathbf{m}_{K-1}\ast e^{-\alpha_{(K-1)i}\mathbf{t}}
\end{bmatrix}.
\end{split}
\end{equation}
Furthermore, the convolution operator in \eqref{eq:Qi_1} can be replaced by a product by Toeplitz matrix defined by the vectors $e^{-\alpha_{ki}\mathbf{t}}$ \citep{Chen2011}, i.e., 
\vspace{-0.1cm}\begin{equation}
\mathbf{Q}_i=\begin{bmatrix}
\mathbf{E}_{1i}\mathbf{m}_1&\cdots&\mathbf{E}_{(K-1)i}\mathbf{m}_{K-1}
\end{bmatrix}.
\end{equation}
with
\vspace{-0.1cm}\begin{equation}
\mathbf{E}_{ki} = T_p(e^{-\alpha_{ki}\mathbf{t}}),
\end{equation}
where $T_p$ is the operator that transforms a vector into a symmetric Toeplitz matrix whose dimensions are defined by the vector length. Note that, $\forall k \in \{1,\ldots,K-1\}$, $\alpha_{k0}=0$  and thus $\mathbf{E}_{k0}=\mathbf{I}_L$ and $\mathbf{Q}_0= [\mathbf{m}_1,\cdots,\mathbf{m}_{K-1}]$.
Also, the matrices of internal coefficients related to the basis functions are given by
\vspace{-0.1cm}\begin{equation}
\begin{split}
\mathbf{B}_i = \begin{bmatrix}
b_{1i1}&b_{1i2}&\cdots&b_{1iN} \\
b_{2i1}&b_{2i2}&\cdots&b_{2iN} \\
\vdots&\vdots&\vdots&\vdots\\
b_{(K-1)i1}&b_{(K-1)i2}&\cdots&b_{(K-1)iN} 
\end{bmatrix},
\end{split}
\end{equation}
with $\underline{\mathbf{B}}=\{ \mathbf{B}_0,\ldots,\mathbf{B}_V \}$.

Besides, additional constraints regarding these sets of parameters are assumed. As in  \eqref{eq:constrM} and \eqref{eq:constrA}, non-negativity is assumed for the factors and corresponding proportions sum to one to reflect physical considerations. Moreover, according to \cite{Gunn2002}, to reduce the indeterminacy of the basis elements solution while allowing for a suitable coverage of the kinetic spectrum, the elements of $\boldsymbol{\alpha}_i = \begin{bmatrix}\alpha_{1i}&\cdots&\alpha_{(K-1)i}\end{bmatrix}$ are assumed to be lower- and upper-bounded by predefined values adjusting the range of expected values. Thus, the kinetic parameter vector $\boldsymbol{\alpha}_i$ ($\forall i\in\{1,\ldots,V\}$) are assumed to belong to the set
\vspace{-0.1cm}\begin{equation}
\begin{aligned}
\boldsymbol{\alpha}_i \in \mathcal{R}\triangleq\{\mathbf{z} \in \mathbb{R}^{K-1}&:\alpha_{i}^{\mathrm{min}}\preceq z_k  \preceq \alpha_{i}^{\mathrm{max}}\}.
\end{aligned}
\label{eq5:alpha_constr}
\end{equation}
A similar choice is adopted for the internal weights, i.e.,
\vspace{-0.1cm}\begin{equation}
\begin{aligned}
\mathbf{B}_i \in \mathcal{B}\triangleq\{\mathbf{z} \in \mathbb{R}^{(K-1) \times N}&:b_{i}^{\mathrm{min}}\preceq z_{kn}  \preceq b_{i}^{\mathrm{max}}\}.
\end{aligned}
\label{eq5:var_constr}
\end{equation}

\subsection{{Corresponding optimization problem}}
\label{sec5:probform}
The PNMM (\ref{eq5:pnmm}) and constraints \eqref{eq:constrM}, \eqref{eq:constrA}, \eqref{eq5:alpha_constr} and \eqref{eq5:var_constr} are combined to formulate a constrained optimization problem. A cost function is thereby defined to estimate the matrices $\mathbf{M}$,  $\mathbf{A}$ and $\boldsymbol{\alpha}$ and the set $\underline{\mathbf{B}}$ containing the matrices $\mathbf{B}_i$. {For the data-fidelity term, we choose the squared Frobenius distance between the dynamic PET image $\mathbf{Y}$ and the proposed data modeling $\mathbf{X}$ defined by PNMM in \eqref{eq5:pnmm}, implicitly assuming Gaussian noise}. Since the problem is ill-posed and nonconvex, additional regularizers become essential. In this work, we propose to define penalization functions $\boldsymbol{\Phi}$, $\boldsymbol{\Psi}$ and $\boldsymbol{\Omega}$ to reflect the available \textit{a priori} knowledge on $\mathbf{M}$, $\mathbf{A}$ and $\underline{\mathbf{B}}$, respectively. The optimization problem is then defined as
\vspace{-0.1cm}\begin{equation}
(\mathbf{M}^*, \mathbf{A}^*, \underline{\mathbf{B}}^*, \boldsymbol{\alpha}^*) \in \argmin_{\mathbf{M}, \mathbf{A}, \underline{\mathbf{B}}, \boldsymbol{\alpha}} \mathcal{J}(\mathbf{M}, \mathbf{A}, \underline{\mathbf{B}}, \boldsymbol{\alpha})  \vspace{-0.10cm}
\label{eq5:optprob}
\end{equation}
with\vspace{-0.10cm}
\vspace{-0.1cm}
\begin{align}
\mathcal{J}&(\mathbf{M}, \mathbf{A}, \underline{\mathbf{B}}, \boldsymbol{\alpha}) =\frac{1}{2} \left\| \mathbf{Y} - \mathbf{M}\mathbf{A}-\sum_{i=0}\mathbf{Q}_i(\mathbf{\tilde{A}}\circ \mathbf{B}_i)\right\|_{\mathrm{F}}^2\\
&+\eta\boldsymbol{\Phi}(\mathbf{A})+\beta\boldsymbol{\Psi}(\mathbf{M})+\lambda\boldsymbol{\Omega}(\underline{\mathbf{B}}) \nonumber\\
&+\imath_{\mathbb{R}_+^{L \times K}}(\mathbf{M})  +\imath_{\mathcal{A}}(\mathbf{A}) +\imath_{\mathcal{R}^{V+1}}(\boldsymbol{\alpha})
+\imath_{\mathcal{B}^{V+1}}(\underline{\mathbf{B}}) \nonumber
\label{eq5:costfct}
\end{align}
where the parameters $\eta$, $\beta$ and $\lambda$ {adjust the regularizations} $\boldsymbol{\Phi}(\mathbf{A})$, $\boldsymbol{\Psi}(\mathbf{M})$ and $\boldsymbol{\Omega}(\underline{\mathbf{B}})$ and $\imath_{\cdot}(\cdot)$ denotes the indicator function on the feasible set associated with the parameter constraints. As proposed by \cite{unmixingCavalcanti2018}, the penalizations $\boldsymbol{\Phi}(\mathbf{A})$ and $\boldsymbol{\Psi}(\mathbf{M})$  associated with factor proportions and factor TACs promote short distances to rough factor TAC estimates and spatially smooth abundance maps, respectively. The penalization function for the variable $\underline{\mathbf{B}}$ is assumed to be separable with respect to (w.r.t.) the tissue compartment, i.e.,
\vspace{-0.1cm}\begin{equation}
\boldsymbol{\Omega}(\underline{\mathbf{B}})=\sum_{i=0}^V \Omega_i(\mathbf{B}_i),
\end{equation}
where $\Omega_i(\mathbf{B}_i)$ promotes spatial sparsity through a group lasso regularizer 
 defined as \citep{factorCavalcanti2018,Ferraris_IEEE_Trans_CI_2017}
 \vspace{-0.1cm}\begin{equation}
\Omega_i(\mathbf{B}_i) = \|\mathbf{B}_i\|_{2,1}=\sum_{n=1}^N{\|\mathbf{b}_{in}\|_{2}}.
\label{eq3:varpen2}
\end{equation}

\subsection{{Optimization with PALM}}
\label{sec5:palm}
As the optimization problem \eqref{eq5:optprob} is nonconvex and nonsmooth, the chosen minimization strategy is the proximal alternating linearized minimization (PALM) scheme \citep{Bolte2013}. It consists in iteratively updating each variable $\mathbf{A}$, $\mathbf{M}$, $\underline{\mathbf{B}}$ and $\boldsymbol{\alpha}$ while all the others are fixed, finally converging to a local critical point $\mathbf{A}^*$, $\mathbf{M}^*$, $\underline{\mathbf{B}}^*$ and $\boldsymbol{\alpha}^*$. The resulting unmixing algorithm, whose main steps are described in the following paragraphs, is summarized in Algo. \ref{algo5:globalplmm}. More details regarding each step are reported in the Appendix.

    \LinesNumbered
    \begin{algorithm}
    \DontPrintSemicolon
    \KwData{$\mathbf{Y}$}
    \KwIn{$\mathbf{A}^{0}$, $\mathbf{M}^{0}$, $\underline{\mathbf{B}}^{0}$, $\boldsymbol{\alpha}^0$}
    
    $\mathbf{A} \leftarrow \mathbf{A}^{0}$\;
    $\mathbf{M} \leftarrow \mathbf{M}^{0}$\;
    $\underline{\mathbf{B}} \leftarrow \underline{\mathbf{B}}^{0}$\;
    $\boldsymbol{\alpha} \leftarrow \boldsymbol{\alpha}^{0}$\;
    \While{stopping criterion not satisfied}{
    \label{algostep5:M} $\mathbf{M} \leftarrow \mathcal{P}_{+}\bigg( \mathbf{M}-\frac{\gamma}{L_M} \nabla_{\mathbf{M}}\mathcal{J}(\mathbf{M}, \mathbf{A}, \underline{\mathbf{B}},\boldsymbol{\alpha})\bigg)$ \;
    \label{algostep5:A} $\mathbf{A} \leftarrow \mathcal{P}_{\mathcal{A}}\bigg( \mathbf{A}-\frac{\gamma}{L_A} \nabla_{\mathbf{A}}\mathcal{J}(\mathbf{M}, \mathbf{A}, \underline{\mathbf{B}},\boldsymbol{\alpha})\bigg)$ \;
 \For{$i\leftarrow 0$ \KwTo $V$}{
    \label{algostep5:B} $\mathbf{B}_i \leftarrow 
    \prox_{\frac{\lambda}{L_{B_i}}\|.\|_1}\bigg(\mathcal{P}_{\mathcal{B}}\bigg(\mathbf{B}_i-\nolinebreak\frac{\gamma}{L_{B_i}} \nabla_{\mathbf{B}_i}\mathcal{J}(\mathbf{M}, \mathbf{A}, \underline{\mathbf{B}},\boldsymbol{\alpha})\bigg)\bigg)$ \;
    }
    \For{$i\leftarrow 1$ \KwTo $V$}{
    \For{$k\leftarrow 1$ \KwTo $K
    $}{
        \label{algostep5:alpha} $\alpha_{ki} \leftarrow 
    \mathcal{P}_{\mathcal{R}}\bigg(\alpha_{ki}-\nolinebreak\frac{\gamma}{L_{\alpha_{ki}}} \nabla_{\alpha_{ki}}\mathcal{J}(\mathbf{M}, \mathbf{A}, \underline{\mathbf{B}},\boldsymbol{\alpha})\bigg)$ \;
    }
    }
    }
    \KwResult{$\mathbf{A}$, $\mathbf{M}$, $\underline{\mathbf{B}}$, $\boldsymbol{\alpha}$}
    \caption{PNMM unmixing: global algorithm \label{algo5:globalplmm}}
    \end{algorithm}

\subsubsection{Optimization with respect to $\mathbf{M}$}
\label{subsec5:optM}

A direct application of the scheme introduced by \cite{Bolte2013} under the constraints defined by \eqref{eq:constrM} leads to the following updating rule
\vspace{-0.1cm}\begin{equation}
\mathbf{M} =\mathcal{P}_{+}\bigg( \mathbf{M}-\frac{\gamma}{L_M} \nabla_{\mathbf{M}}\mathcal{J}(\mathbf{M}, \mathbf{A}, \underline{\mathbf{B}},\boldsymbol{\alpha})\bigg)
\label{eq5:Mpalm}
\end{equation}
where $\mathcal{P}_{+}(\cdot)$ is the projector onto the nonnegative set $\{\mathbf{X}|\mathbf{X}\succeq \mathbf{0}_{L,R}\}$. Moreover, $L_{m_{k}}$ is a bound on the Lipschitz constant of $\nabla_{\mathbf{\tilde{M}}}\mathcal{J}(\mathbf{m}_k, \mathbf{A}_k,\mathbf{W}_k,\mathbf{E}_k)$.

\subsubsection{Optimization with respect to $\mathbf{A}$}
\label{subsec5:optA}
Similarly to paragraph \ref{subsec5:optM}, the abundance update is defined as the following
\vspace{-0.1cm}\begin{equation}
\mathbf{A} = \mathcal{P}_{\mathcal{A}}\bigg( \mathbf{A}-\frac{\gamma}{L_A} \nabla_{\mathbf{A}}\mathcal{J}(\mathbf{M}, \mathbf{A}, \underline{\mathbf{B}},\boldsymbol{\alpha})\bigg)
\label{eq5:Apalm}
\end{equation}
where $\mathcal{P}_{\mathcal{A}}(\cdot)$ is the projection on the set $\mathcal{A}$ defined by the abundance constraints (\ref{eq:constrA}), which can be computed with efficient algorithms (see, e.g., the strategies discussed by \cite{Condat2015}). Likewise, $L_{\tilde{A}}$ is the Lipschitz constant of $\nabla_{\tilde{A}}\mathcal{J}(\mathbf{\tilde{M}}, \mathbf{\tilde{A}}, \mathbf{Q},\underline{\mathbf{B}})$.

\subsubsection{Optimization with respect to $\mathbf{B}_i$}
\label{subsec5:optbeta}

The updating rule for the basis function coefficients, under the constraints defined by (\ref{eq5:var_constr}),  can be written as
\begin{equation*}
\mathbf{B}_i = 
    \prox_{\frac{\lambda}{L_{B_i}}\|.\|_1}\bigg(\mathcal{P}_{\mathcal{B}}\bigg(\mathbf{B}_i-\nolinebreak\frac{\gamma}{L_{B_i}} \nabla_{\mathbf{B}_i}\mathcal{J}(\mathbf{M}, \mathbf{A}, \underline{\mathbf{B}},\boldsymbol{\alpha})\bigg)\bigg),
\label{eq5:Bpalm}
\end{equation*}
where $\mathcal{P}_{\mathcal{B}}$ is the projection into the set $\mathcal{B}$ defined by \eqref{eq5:var_constr}.
Indeed, the proximal map of the sum of an indicator function and the $\ell_1$-norm is exactly the composition of the proximal maps of both individual functions, following the same principle shown by \cite{Bolte2013}.
$L_{B_i}$ is the Lipschitz constant of $\nabla_{B_i}\mathcal{J}(\mathbf{\tilde{A}}, \mathbf{B},\mathbf{Q})$.

\subsubsection{Optimization with respect to $\alpha_{ki}$}
\label{subsec5:optalpha}
Finally, the updating rule for the basis function exponential coefficients, under the constraints in (\ref{eq5:alpha_constr}), is 
\vspace{-0.1cm}\begin{equation}
\alpha_{ki} = 
    \mathcal{P}_{\mathcal{R}}\bigg(\alpha_{ki}-\nolinebreak\frac{\gamma}{L_{\alpha_{ki}}} \nabla_{\alpha_{ki}}\mathcal{J}(\mathbf{M}, \mathbf{A}, \underline{\mathbf{B}},\boldsymbol{\alpha})\bigg).
\end{equation}
Also, $\mathcal{P}_{\mathcal{R}}$ is the projection into the set $\mathcal{R}$ defined in \eqref{eq5:alpha_constr}.
The Lipschitz constant is $L_{\alpha_{ki}}$.
\section{Evaluation on synthetic data}
\label{sec5:synt_results}
\subsection{Synthetic data generation}
To illustrate the accuracy of our algorithm, experiments are conducted on one $128 \times 128\times 64$-pixel synthetic image with $L=27$ times of acquisition ranging from $1$ to $15$ minutes for a total period of $90$ minutes. In this image, each voxel is constructed as a combination of $K=3$ pure classes representative of the brain, which is the organ of interest in the present work: pure nSB gray matter, pure nSB white matter and pure blood or veins. {Moreover, SB TACs are a result of nonlinearities affecting the pure nSB factors and, therefore, they do not represent new factors.}
The image is generated from the high resolution dynamic PET numerical phantom by \cite{zubal1994computerized} with TACs generated from real PET images acquired with the Siemens HRRT and injected with $^{18}$F-DPA. The overall generation process is described in what follows:

\begin{itemize}
\item The dynamic PET phantom has been first linearly unmixed using the N-FINDR \citep{Winter1999spie} and SUnSAL \citep{Bioucas2010} algorithms with an initial number of classes of 4, accounting for SB and nSB gray matter, white matter and blood. The TAC factor for SB gray matter found by N-FINDR is discarded while the other factors are selected to constitute the ground-truth non-specific factor TACs $\mathbf{m}_1,...,\mathbf{m}_K$. The {4} factor proportions found by SUnSAL are used to generate {4} binary maps after a thresholding. 
\item The binary maps of SB and nSB gray matter generated from the SUnSAL output are merged to yield a general gray matter factor proportion. The white matter and blood binary maps are directly used as factor proportions. {The final 3 binary maps are shown in the second row of Fig. \ref{fig5:A_cnu}.}
\item The SB gray matter binary map is used to provide the location of the weight coefficients of nonlinearity in the gray matter. An anomaly binary map is generated inside the white matter factor proportion to provide the location of SB in white matter.
\item The weights and exponential coefficients describing the nonlinearities are generated from the two-tissue full reference model by \cite{Haggstrom2016}. Moreover, two overall levels of binding are generated for each tissue by altering the tracer delivery ratios $R_{1n}$ ($n = 1,\ldots, N$) between the tissue of interest and the reference tissue. More precisely, for the SB gray matter, the tracer delivery ratios $R_{1n}$ take values in $\left\{1,1.6\right\}$ according to a spatially coherent pattern{, i.e., the same $R_{1n}$ is set to each pixel inside a non-zero uniform region of the binary map.} The other parameters are set to $k_2=0.4$, $k_3=0.15$ and $k_4=0.01$ for all non-zero pixels of the binary map. For the SB white matter, similarly {the  parameters are set to} $R_{1n} \in \left\{1,1.6\right\}$, $k_2=0.3$, $k_3=0.15$ and $k_4=0.01$. 
\item Parameter generation directly yields the exponential coefficients $\boldsymbol{\alpha}_1, \cdots, \boldsymbol{\alpha}_V$, {while the computed weights are inserted in place of the non-zero values of the previously defined SB binary maps for the gray and white matters, accordingly}.
\end{itemize}

After the phantom generation process, a PSF defined as a space-invariant and isotropic Gaussian filter with $\textrm{FWHM} = 4.4$mm is applied to the output image.  Finally, as proposed by \cite{unmixingCavalcanti2018}, the measurements are corrupted by a Gaussian noise with a signal-to-noise ratio (SNR) of $20$dB.  { Simulations were conducted with $20$ different noise realizations to get statistically reliable performance measures.} 
For {sake of} fair comparison, the ground-truth for the factor proportions $\mathbf{A}$ {(first row of Fig. \ref{fig5:A_cnu})} and the basis function coefficients $\underline{\mathbf{B}}$ corresponding to nonlinearities {are computed by applying the proposed algorithm with fixed factors and exponential coefficients, i.e., by updating only the two variables of interest, in an image with no noise affected by the PSF. This allows the partial volume effect to be taken into account.}

\begin{table}
\centering
\caption{Model parameters}
\begin{tabular}{|c|c|c|}
\hline
 & PNMM&SLMM \\
\hline
$\eta$ & 0.500& 0.500 \\
\hline
$\beta$ & 0.100& 0.100 \\
\hline
$\lambda$ & 0.500& 0.500 \\
\hline
\end{tabular}
\label{table5:par_pnmm}
\end{table}

\subsection{Compared methods}
The {proposed PNMM unmixing technique} is compared against the basis pursuit method by \cite{Gunn2002}, referred to as DEPICT, and the specific binding linear mixing model (SLMM) unmixing {we already} proposed \citep{unmixingCavalcanti2018}. Their implementations are detailed in what follows.

\noindent \textbf{DEPICT --}
In this work, DEPICT is implemented with proximal gradient steps for comparison purposes. As proposed by \cite{Gunn2002}, basis pursuit denoising is conducted with 30 basis functions logarithmically spaced between 0.03 and 6 min$^{-1}$ and an additional basis function to represent the offset. The number of basis functions is fixed to $31$ as a trade-off between precision and computation time. For comparison purposes, DEPICT is conducted with two different reference TACs: first the gray, then the white matter factors {that are defined as described in the following PNMM-unmixing dedicated paragraph (see below).}

\noindent \textbf{SLMM-unmixing --}
To appreciate the interest of extracting a physically interpretable quantity, the proposed algorithm is also compared with the previous method introduced by \cite{unmixingCavalcanti2018}. {This SLMM method considers an additional class dedicated to specific binding}. The penalizations chosen for $\mathbf{M}$ and $\mathbf{A}$ are the same, as the one for matrix $\mathbf{B}$ in SLMM and the set of matrices $\underline{\mathbf{B}}$ in PNMM. Thus, we consider the same regularization parameters (see Table \ref{table5:par_pnmm}) for fair comparisons. The algorithm is stopped when the normalized difference between two consecutive objective function values is below a threshold  $\varepsilon$ set to $5 \times 10^{-3}$. The variability dictionary is learned from a predefined high-uptake region of the image, comprising both SB gray and white matters. {Blood, white matter and gray matter }factors and their corresponding proportions are initialized as described in the next paragraph dedicated to PNMM (see below). {The factor and factor proportion related to specific binding are initialized with zeros, as is the matrix $\mathbf{B}$}. We allow the method to run $50$ iterations with fixed $\mathbf{M}$ so as to improve the initializations of $\mathbf{A}$ and $\mathbf{B}$, while preventing factors from merging.

\noindent \textbf{PNMM-unmixing --}
{For the proposed algorithm}, factor proportions are initialized with the binary maps coming from the generation process. In a real image, this is equivalent to using an MRI segmentation to produce the maps of tissues. Regarding the initialization of the factors, the TACs from each tissue are organized from the lowest area-under the curve (AUC) TAC to the highest. The first 10\% AUC TACs are discarded and we average the TACs whose AUC are the 10\% to 20\% lowest ones. Then, the basis functions and their corresponding coefficients are computed with an instance of our algorithm, where factors and proportion maps are not updated. As suggested by \cite{Gunn2002}, the exponential coefficients for our method will be bounded with $\alpha_{i}^{\mathrm{min}}=d_c$ ($\forall i$), where $d_c=0.0063$  min$^{-1}$ is the decay constant for $[18F]$, and $\alpha_{i}^{\mathrm{max}}=6$ min$^{-1}$. Limits are also imposed for the nonlinearity coefficients. We know that $b_{0n}={R}_{1{n}}-1$ cannot be nonnegative in SB tissues when nSB tissues are used as a reference. Moreover, the maximum value of $\mathrm{R}_{1{n}}$ is generally not higher than $1.7$. 
Thus, we choose $b_{0}^{\mathrm{min}}=0$ and $b_{0}^{\mathrm{max}}=0.7$. The limits for the nonlinearity coefficients can be chosen by analyzing the relations of the known kinetic parameters of the tracer under study and the weights. In our study, we found it was sufficient to choose $b_{1}^{\mathrm{min}}=-0.2$, $b_{1}^{\mathrm{max}}=0$, $b_{2}^{\mathrm{min}}=0$ and $b_{2}^{\mathrm{max}}=0.15$. Finally, Table \ref{table5:par_pnmm} reports the values of the regularization parameters.

\subsection{Figures-of-merit}
The performance of the method is measured by computing the normalized mean square error (NMSE)
\vspace{-0.1cm}\begin{equation}
 \mathrm{NMSE}(\hat{\boldsymbol{\theta}}_i)=\frac{ \| \hat{\boldsymbol{\theta}}_i-\boldsymbol{\theta}_i^*\|_{\mathrm{F}}^2}{ \|\boldsymbol{\theta}_i^*\|_{\mathrm{F}}^2},
\end{equation}
where $\boldsymbol{\theta}_i^*$ and $\hat{\boldsymbol{\theta}}_i$ are the {true} and estimated latent variables, respectively. The following variables are evaluated with this metric: the factor proportions $\mathbf{A}$, the non-specific factor TACs $\mathbf{M}$, the tracer delivery ratio $\mathbf{R}_1=\left[R_{11},\ldots,R_{1N}\right]^T$, the kinetic parameter $\boldsymbol{\alpha}$ and the binding potential w.r.t. the free fractions in tissue $\mathbf{BP}.f_{\mathrm{T}}$.

\subsection{Results and discussion}
\label{sec5:results}
\begin{figure}
\begin{center}
\includegraphics[width=\columnwidth]{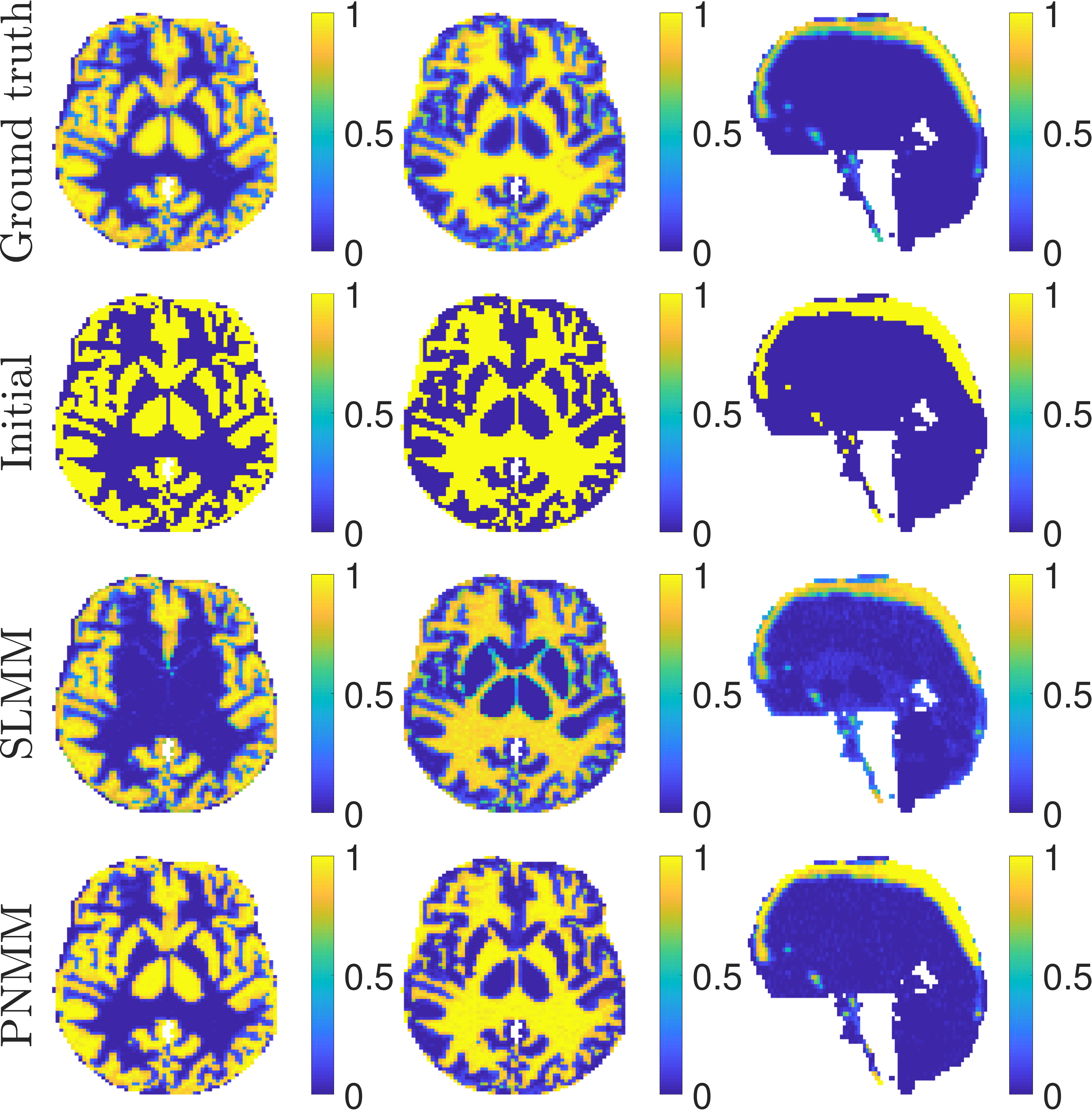}
\caption{Factor proportion maps obtained from the synthetic image corresponding to the gray matter, white matter and blood, from left to right. The first 2 columns show transaxial views while the last one shows a sagittal view. All images are in the same scale $[0,1]$.}
\label{fig5:A_cnu}
\end{center}
\end{figure}

Fig. \ref{fig5:A_cnu} shows, from top to bottom: the ground-truth factor proportion, the initial segmentation and the final SLMM and PNMM results. The first column shows the gray matter, the second column shows the white matter and the third column presents the blood factor proportion. Visual analysis suggests that both SLMM and PNMM techniques are able to include the partial volume effect into the factor proportions, as expected. 
{Note that the SB gray and white matters do not appear on the corresponding SLMM factor proportion, since the algorithm deals with nSB and SB as different tissues.} 

\begin{figure*}
\centering
\includegraphics[width=0.7\textwidth]{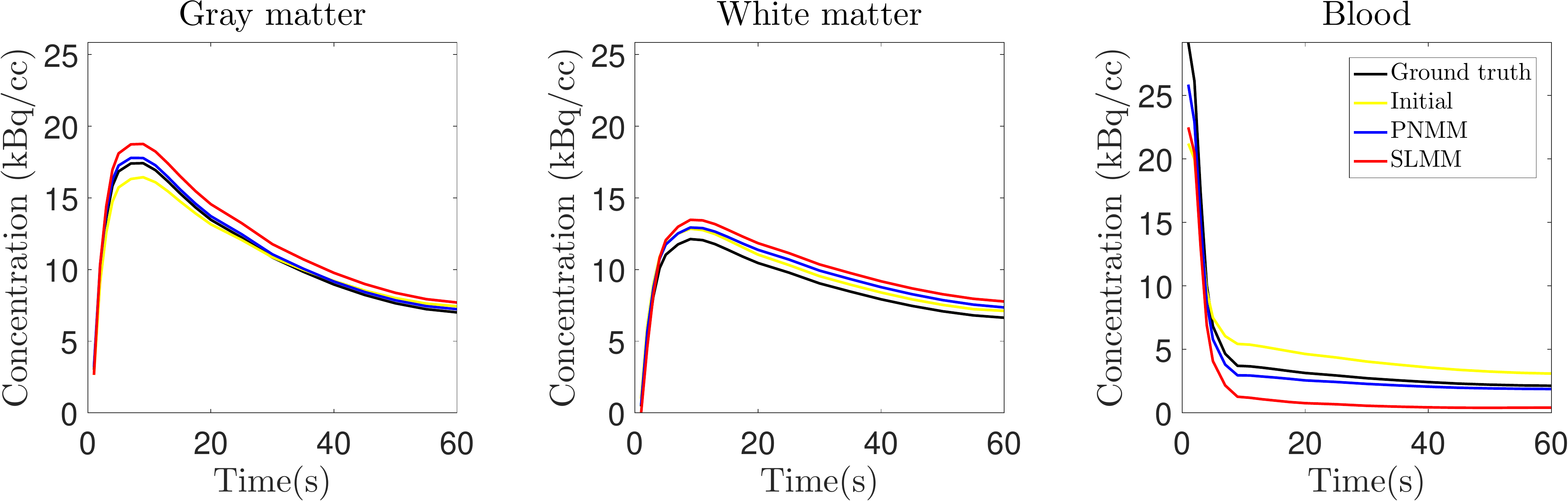}
\caption{Factor TACs estimated from the synthetic image.}
\label{fig5:M_cnu}
\end{figure*}

{The estimated SB gray and white matter factors are displayed in Fig. \ref{fig5:M_cnu}, together with their initialization.} Visual comparison suggests that PNMM improves the initial factor estimation with final global TACs that are very close to the ground-truth. {Comparatively, the SLMM result for the blood TAC is clearly less accurate.} 

These results are further confirmed by the quantitative evaluation of Table \ref{table5:nmse} that shows the NMSE of the variables of interest as {estimated} during the initialization and after conducting SLMM and PNMM unmixing {for the 20 noise realizations}. {As SLMM is supposed to identify the SB regions with an exclusive factor, it would be quite unfair to crudely interpret that its estimated abundance maps is poor, since this initialization framework is not ideal for SLMM.} 
The PNMM results for both $\mathbf{A}$ and $\mathbf{M}$ are remarkably improved. This is a favorable outcome, since it suggests that PNMM is able to improve the results with this initialization setting that can be easily replicated in real image applications.

\begin{table}
\centering
\caption{NMSE of $\mathbf{A}$, $\mathbf{M}$ and $\mathbf{BP}$ as chosen in initialization and after conducting PNMM-unmixing.}
\begin{tabular}{|c|c|c|c|}
\hline
 & Initial & PNMM &SLMM\\
\hline
$\mathbf{A}$          & $0.236 $                    & $0.101$ & $0.421$ \\
&&\footnotesize{$\pm 1\times10^{-5}$} & \footnotesize{$ \pm 4\times10^{-6}$} \\
\hline
$\mathbf{M}$          & $0.165$ & $0.136$ & $0.182$ \\
&\footnotesize{$\pm 4 \times 10^{-8}$} & \footnotesize{$ \pm 5\times10^{-7}$}& \footnotesize{$\pm 5\times10^{-7}$} \\
\hline
$\mathbf{R}_1$        & $1.304$ & $0.601 $&- \\
&\footnotesize{$\pm 1 \times 10^{-4}$} & \footnotesize{$\pm 1\times10^{-5}$}& \\
\hline
$\boldsymbol{\alpha}$ & $0.210$ & $0.204 $&- \\
&\footnotesize{$\pm 6 \times 10^{-10}$} & \footnotesize{$\pm 5\times10^{-8}$}&\\
\hline
$\mathbf{BP}.f_{\mathrm{T}}$     & $0.156$ & $0.097$&- \\
&\footnotesize{$\pm 1 \times 10^{-7}$} & \footnotesize{$\pm 8\times10^{-8}$}&\\
\hline
\end{tabular}
\label{table5:nmse}
\vspace{-0.5cm}
\end{table}

\begin{figure}[h]
\begin{center}
\includegraphics[width=0.7\columnwidth]{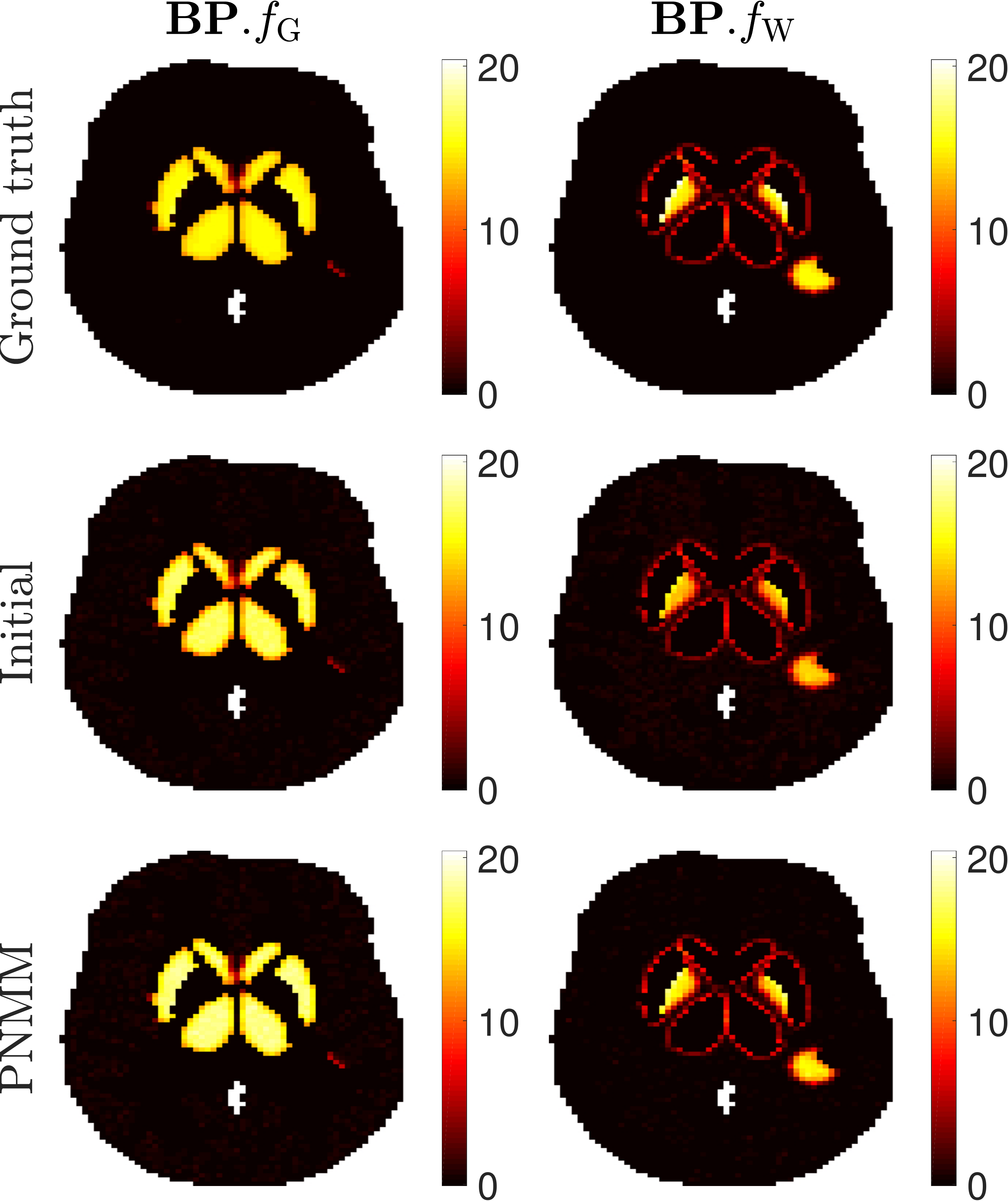}
\caption{From top to bottom: ground-truth, initial and PNMM estimations of $\mathbf{BP}.f_{\mathrm{T}}$. The first column corresponds to the gray matter and the second to the white matter.}
\label{fig5:B_cnu}
\end{center}
\vspace{-0.5cm}
\end{figure}

Fig. \ref{fig5:B_cnu} shows the binding potential w.r.t. the free fractions in tissue $\mathbf{BP}.f_{\mathrm{T}}$ for the gray matter (left) and white matter (right). The first two rows present the ground-truth and initial $\mathbf{BP}.f_{\mathrm{T}}$ and the last row presents the PNMM estimation of $\mathbf{BP}.f_{\mathrm{T}}$ in the PNMM formulation, where there are two $\mathbf{BP}.f_{\mathrm{T}}$ to be estimated in the same setting: one for the gray matter ($\mathbf{BP}.f_{\mathrm{G}}$) and one for the white matter ($\mathbf{BP}.f_{\mathrm{W}}$). It is hard to determine by visual comparison whether the binding potential is improved from initialization by the PNMM-unmixing. Table \ref{table5:nmse} presents the quantitative results of the NMSE for the matrix $\mathbf{R}_1$, where the estimated $\mathbf{b}_{0n}$ includes the coefficients for both non-specific tissues, and the NMSE for the matrix $\mathbf{BP}.f_{\mathrm{T}}$ with the binding potential in each voxel for each tissue. Quantitative results suggest that $\mathbf{BP}.f_{\mathrm{T}}$ is better estimated by conducting the whole PNMM-unmixing, which is natural since the estimations of the factors and factor proportions are also improved. Moreover, the ratio of delivery of the tracer $\mathbf{R}_1$ seems to show a much greater improvement. 

\begin{figure}
\begin{center}
\includegraphics[width=0.6\columnwidth]{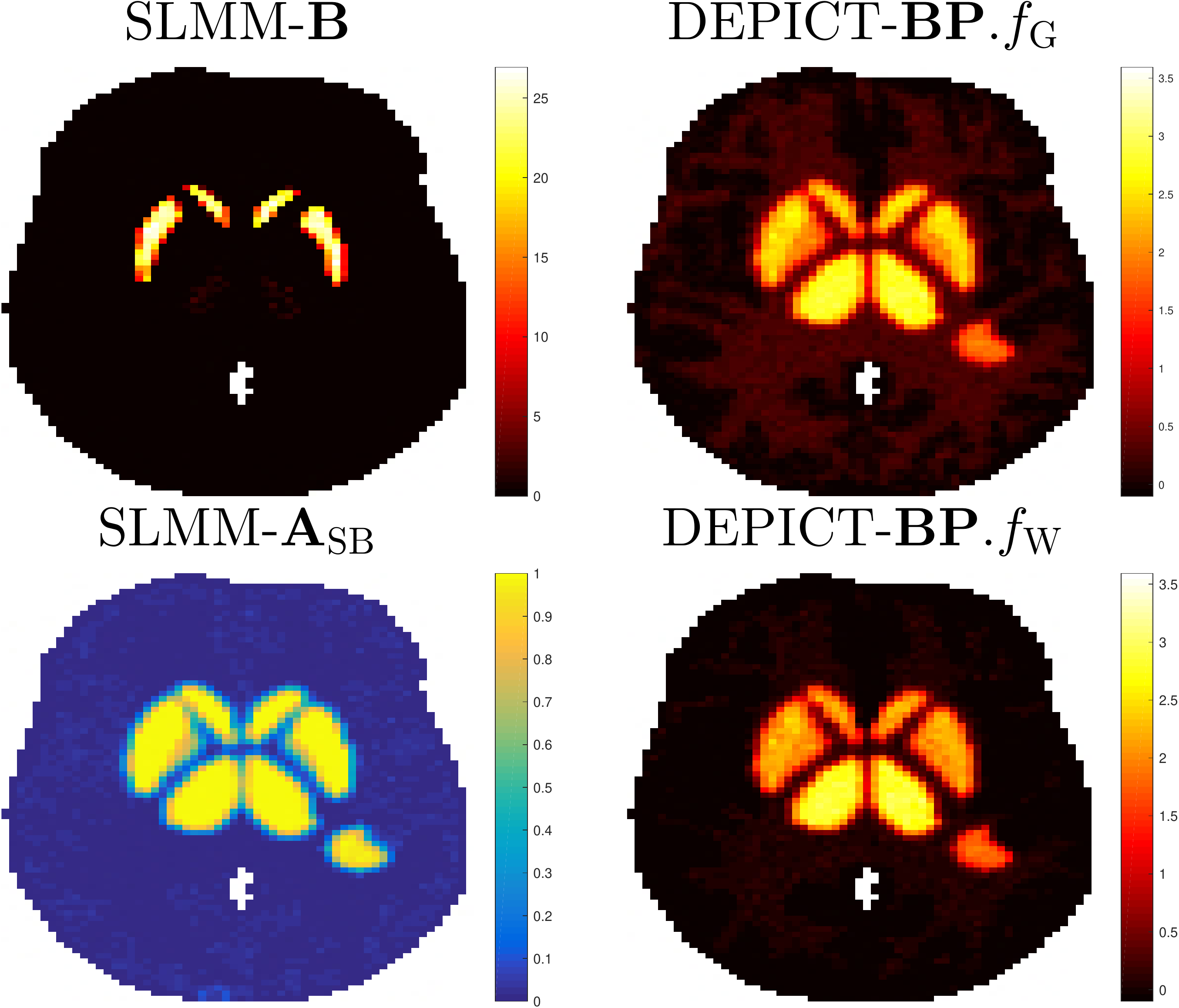}
\caption{The left column shows SLMM variability result (top) and SLMM factor proportion related to the SBF (bottom) and the right column shows DEPICT $\mathrm{BP}.f_{\mathrm{T}}$ estimation using the gray (top) and white (bottom) matters as reference TACs, respectively. {Note that DEPICT $\mathbf{BP}.f_{\mathrm{T}}$ was estimated for the whole image using the respective tissue TAC as reference.}}
\label{fig5:B_depict_slmm}
\end{center}
\vspace{-0.5cm}
\end{figure}

{Moreover, for comparison, Fig. \ref{fig5:B_depict_slmm} reports the SLMM results for the factor proportion and the internal variability (left column) and the DEPICT results taking the white matter as reference TAC and the gray matter as reference TAC (right column).} SLMM results are not equivalent to the binding potential or any other physical quantity of clinical use. Still, it is possible to see that SB tissues have been identified and the missing regions from gray and white matter factor proportions of Fig. \ref{fig5:A_cnu} are relocated in the SB factor proportion. The evaluation of $\mathbf{R}_1$ and $\mathbf{BP}.f_{\mathrm{T}}$ cannot be conducted for the DEPICT result, since the ground-truths are not equivalent. Visual inspection suggests that DEPICT is able to correctly locate the specific binding tissues with similar intensities of $\mathbf{BP}.f_{\mathrm{T}}$. The gray matter result presents some binding in the white matter tissue, showing the potential bias that could be expected when the whole image is represented by one reference TAC, while considering distinct reference TACs in distinct non-specific binding tissues {seems} more accurate. Even though the overall result may often be sufficient for clinical applications, given the challenge of interpreting dynamic PET images, they seem to be less accurate than the method herein proposed, in terms of both BP intensities and location. Moreover, DEPICT does not allow the user to differentiate the tissue that is affected, for instance, by an abnormality, while our method may provide this detailed information.

Note that if the factor proportions are initialized with an MRI segmentation and fixed, PNMM works as a local reference model, where each non-specific tissue of the image is treated as a different region-of-interest (ROI) and is therefore allowed to have its own reference TAC. This is equivalent of conducting DEPICT in each segmented tissue, but allowing the global reference TAC to be improved in each step. This setting is also able to provide the tissues affected by specific binding but does not take into account the partial volume effect.

\section{Evaluation on real data}
\label{sec5:real_results}
\subsection{PET data acquisition}
To assess the performance of the proposed approach on a real dataset, the compared methods have been applied to a dynamic PET image acquired with an Ingenuity TOF Camera from Philips Medical Systems of a stroke subject injected with [18F]DPA-714, seven days after the stroke. The PET acquisition was reconstructed into a $128 \times 128\times 90$-voxels dynamic PET image with $L=31$ time-frames. The PET scan image registration time ranged from $10$ seconds to $5$ minutes over a $59$ minutes period. The voxel size was of $2 \times 2 \times 2$ mm$^3$.  

\subsection{Compared methods}
Factor proportions are initialized with binary maps mainly constituted from a manually labelled MRI segmentation and improved with a K-means result for the voxels that were not labelled in the MRI segmentation. Factors are initialized as in the synthetic case. 
The stroke region is segmented on this registered MRI image. It is used to define a set of voxels used to learn the spatial variability descriptors for SLMM-unmixing. The nominal SBF for SLMM is fixed as the empirical average of the corresponding TACs with AUC comprised between the $10$th and $20$th percentile. The matrix $\mathbf{B}$ is initialized with zeros and, as before, we allow the method to run 50 iterations with fixed $\mathbf{M}$.

For PNMM, the basis functions and their corresponding coefficients are initialized as in the synthetic case, with an instance of our algorithm, where factors and proportion maps are not updated. The exponential coefficients are constrained according to \eqref{eq5:alpha_constr} with $\alpha_{\mathrm{min}}=0.034$  min$^{-1}$ and $\alpha_{\mathrm{max}}=6$ min$^{-1}$. The nonlinearity coefficients are constrained following \eqref{eq5:var_constr} with $b_{i}^{\mathrm{min}}=-1$ and $b_{i}^{\mathrm{max}}=1$ ($i=0,\ldots,V$). 

\subsection{Results and discussion}
\begin{figure*}
\centering
\includegraphics[width=\textwidth]{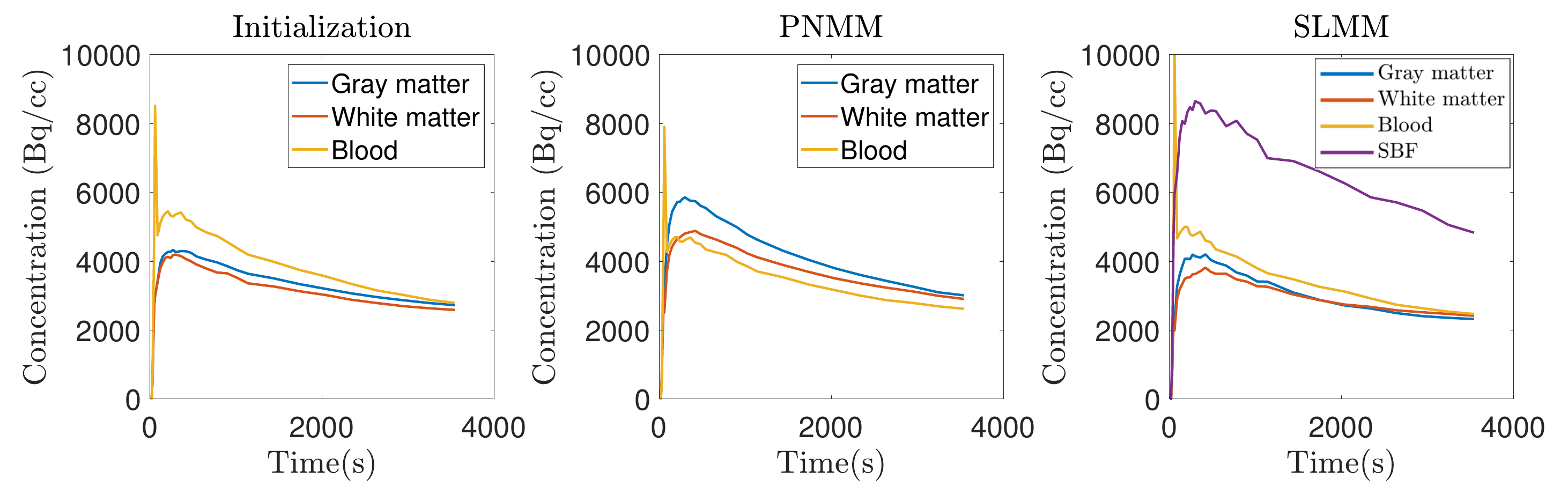}
\caption{Factor TACs estimated from the real image}
\label{fig5:M_real}
\end{figure*}

 Fig. \ref{fig5:M_real} presents the initial and estimated factor TACs. 
 The initial gray and white matter factor TACs are very similar. SLMM estimates gray and white matter TACs that are lower than initialization. This may be because the SBF, which is fixed, is very close to the gray matter factor TAC, inducing the gray matter factor to be smaller. On the other hand, PNMM is able to differentiate the gray and white matter factor TACs in both intensity and shape. {It is also able to increase the AUC of both white and gray matter factors so it is higher than the blood factor AUC, which is expected since the blood is supposed to present a peak of concentration at the moment of radiotracer injection and then reduce the concentration over time.} 

\begin{figure}[H]
\begin{center}
\includegraphics[width=0.8\columnwidth]{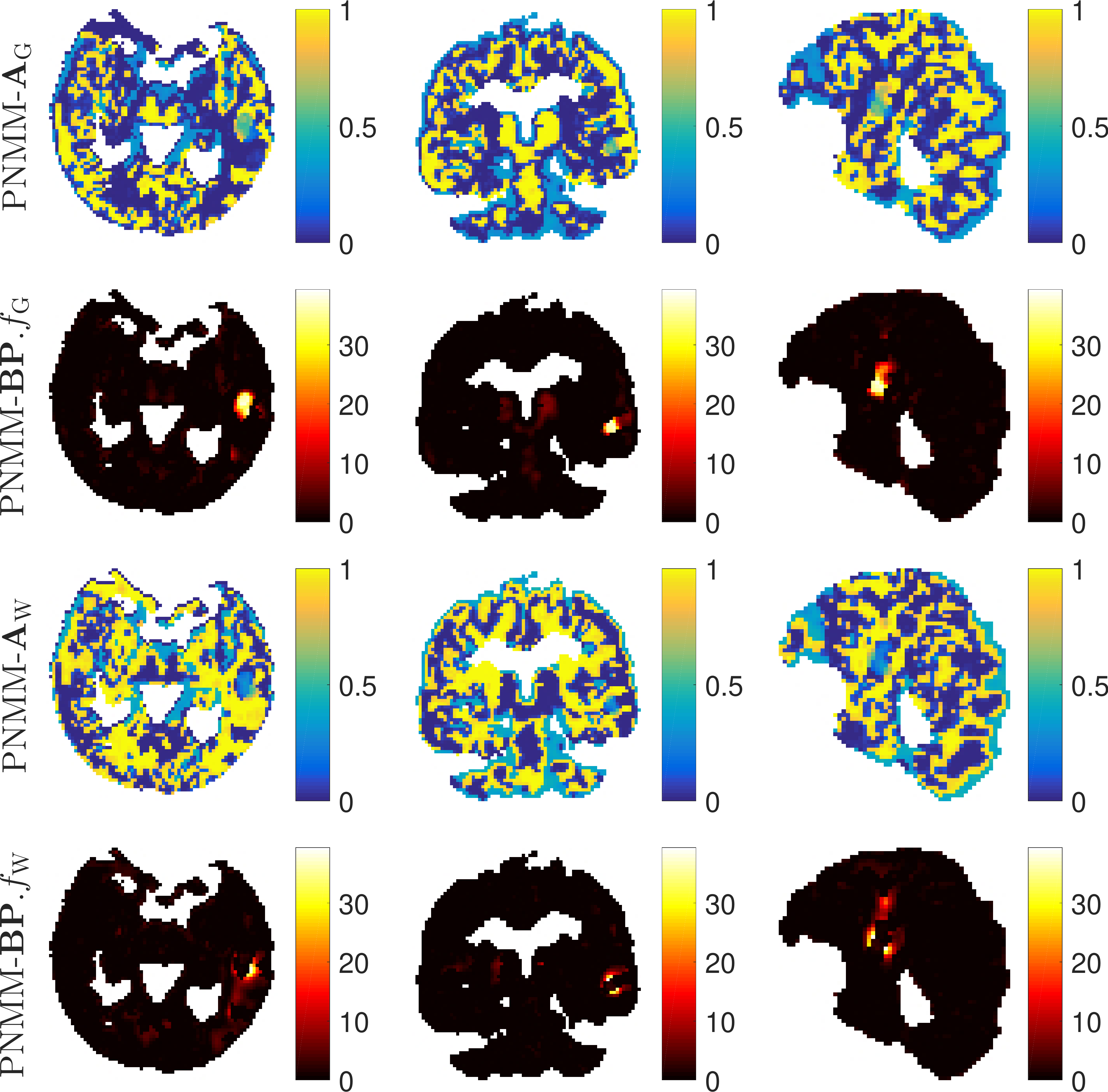}
\caption{ From top to bottom: gray matter factor proportion, $\mathbf{BP}.f_{\mathrm{T}}$ for the gray matter (in $(\text{mL plasma})\cdot (\text{mL tissue})^{-1} $), white matter factor proportion and $\mathbf{BP}.f_{\mathrm{T}}$ for the gray matter (in $(\text{mL plasma})\cdot (\text{mL tissue})^{-1} $), all from PNMM results. From left to right: transaxial, coronal and sagittal view.}
\label{fig5:3D_SBR_real_T}
\end{center}
\vspace{-0.5cm}
\end{figure}

{Fig. \ref{fig5:3D_SBR_real_T} shows a 3D visualization of the PNMM results corresponding to the different tissues, i.e., from bottom to top, the gray matter factor proportion $\mathbf{A}_{\mathrm{G}}$, the gray matter BP, i.e., $\mathbf{BP}.f_{\mathrm{G}}$, the white matter factor proportion $\mathbf{A}_{\mathrm{W}}$ and the white matter BP, i.e., $\mathbf{BP}.f_{\mathrm{W}}$. The gray and white matter binding potentials as well as factor proportions seem complementary, i.e., the SB regions missing in the PNMM binding potential for the gray matter can be found in the result for the white matter and the same is valid for the factor proportions. Note that the $\mathbf{BP}$ for each tissue matches the corresponding factor proportion as a consequence of the mathematical formulation of the problem. This result highlights the fact that the algorithm is not only able to correctly locate the region affected by the stroke, but is also able to identify the affected tissues.}

\begin{figure}[H]
\begin{center}
\includegraphics[width=0.85\columnwidth]{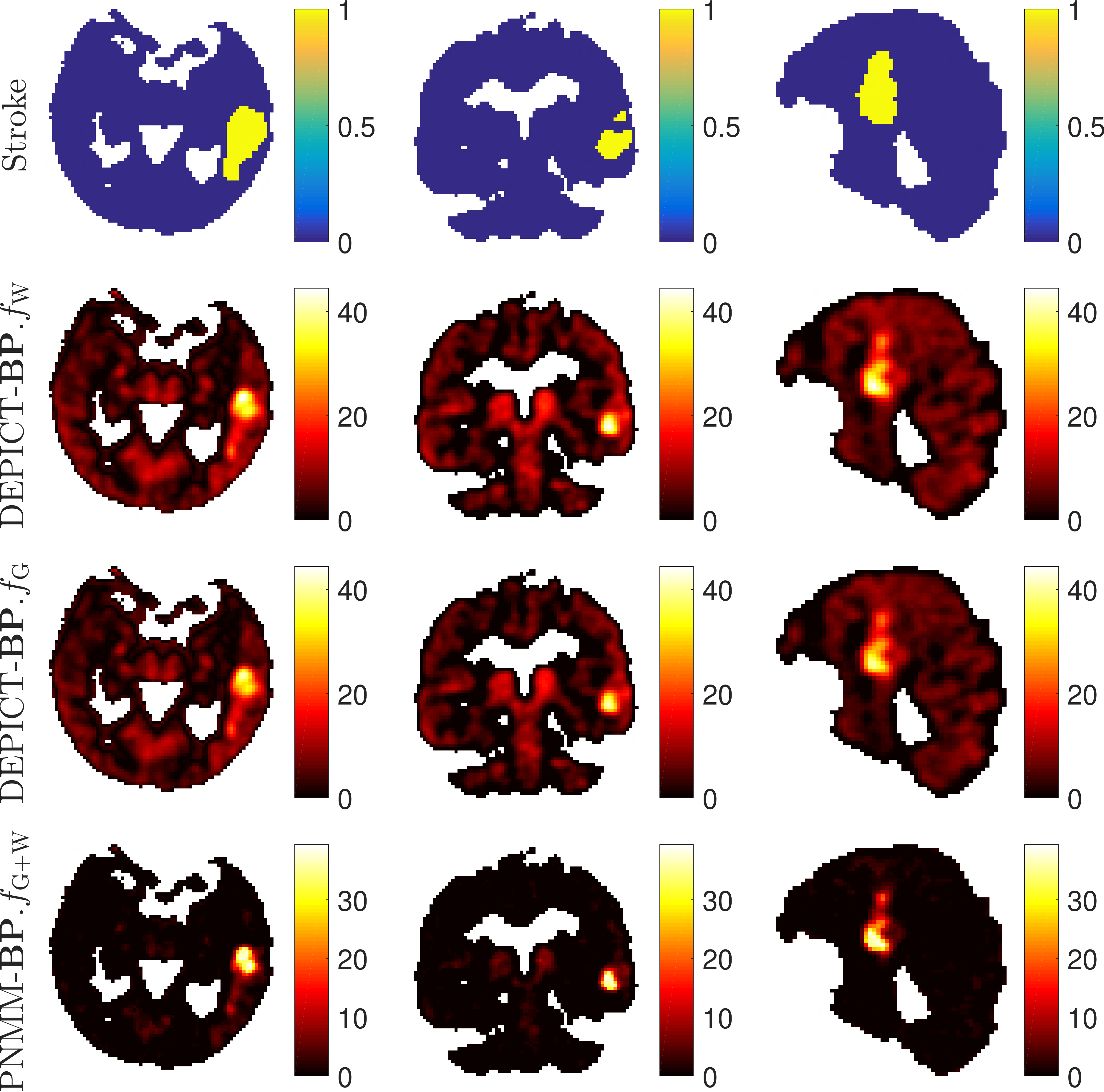}
\caption{From top to bottom: stroke region, DEPICT $\mathbf{BP}.f_{\mathrm{T}}$ (in ${{({\text{mL plasma}})}/{({\text{mL tissue}})}} $) 
   using the white matter TAC as reference, DEPICT $\mathbf{BP}.f_{\mathrm{T}}$ (in ${{({\text{mL plasma}})}/{({\text{mL tissue}})}} $) 
  using the gray matter TAC as reference and total $\mathbf{BP}.f_{\mathrm{T}}$ (in ${{({\text{mL plasma}})}/{({\text{mL tissue}})}} $) 
  estimated by PNMM on both gray and white matter. From left to right: transaxial, coronal and sagittal view.}
\label{fig5:3D_SBR_real_all}
\end{center}
\end{figure}

{Fig. \ref{fig5:3D_SBR_real_all} shows a 3D visualization of the specific binding region, i.e., from top to bottom, the stroke segmented with an MRI, the DEPICT $\mathbf{BP}.f_{\mathrm{W}}$ result using the initial white matter TAC as reference, the DEPICT $\mathbf{BP}.f_{\mathrm{G}}$ result using the initial gray matter TAC as reference and the PNMM $\mathbf{BP}.f_{\mathrm{G}+\mathrm{W}}$ result corresponding to the maximum pixel value between $\mathbf{BP}.f_{\mathrm{G}}$ and $\mathbf{BP}.f_{\mathrm{W}}$, so as to provide a complete representation of the stroke region. Visual inspection suggests that DEPICT presents high $\mathbf{BP}$ intensities in brain regions not expected to be affected by SB. In comparison, PNMM seems  visually more accurate, capturing with precision the stroke region. This is due to the simultaneous update of the factors, that improves the initialization result, as seen in Fig. \ref{fig5:M_real}.}. 

{While DEPICT provides the binding potential for the entire image, the PNMM result is able to provide the SB voxels for each tissue. In addition, by combining the results per tissue, we are able to acquire a complete vision of the binding potential. To summarize, it seems that PNMM is able to locate the specific binding in the entire image as soon as we combine the results for the different tissues. Compared to DEPICT the results seem more accurate, partly because using the same reference TAC in DEPICT is expected to bias the results.}

\begin{figure}[h]
\begin{center}
\includegraphics[width=0.9\columnwidth]{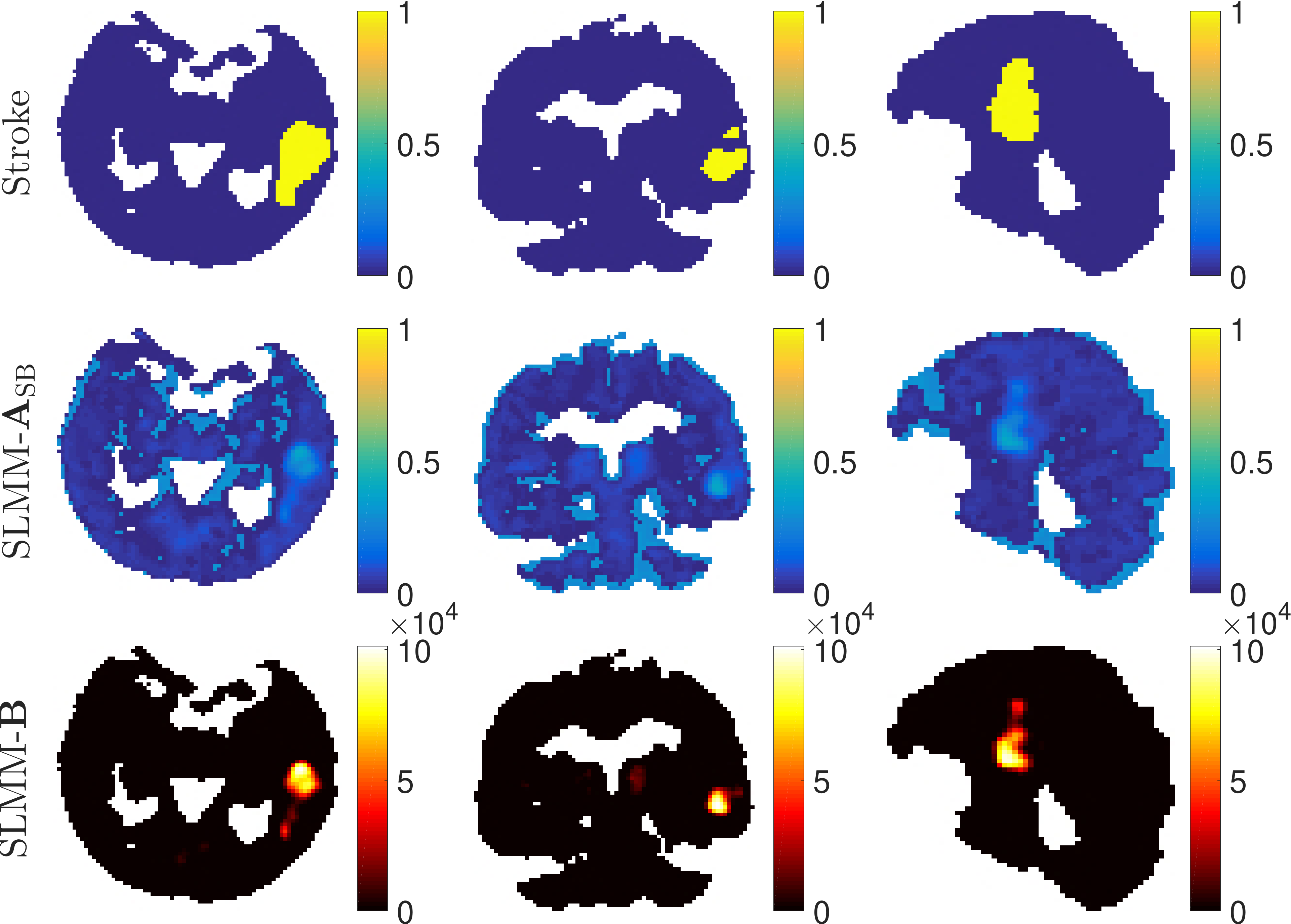}
\caption{From top to bottom: stroke region, SB factor proportion estimated by SLMM and internal variability estimated by SLMM. From left to right: transaxial, coronal and sagittal view.}
\label{fig5:3D_SBR_real_SLMM}
\end{center}
\end{figure}

Finally, Fig. \ref{fig5:3D_SBR_real_SLMM} shows the SLMM $\mathbf{A}_{\mathrm{SB}}$ factor proportion and internal variability $\mathbf{B}$ results in this setting. {The $\mathbf{A}_{\mathrm{SB}}$ result presents a soft highlight on the SB area affected by the variability, which is a relevant outcome, especially because this factor proportion was initialized with zeros. The variability $\mathbf{B}$, that has no physically meaningful unit, correctly identifies the SB area, which is complementary to the information brought by the corresponding factor proportion.} 
Although the result is informative, it is not complete in terms of clinical assessment, in opposition to the other two methods studied.

\section{Conclusion}
\label{sec5:conclu}
This paper presented a novel technique for dynamic PET analysis that combined nonlinear unmixing and parametric imaging to yield a clinically interpretable result for factor analysis. To this end, this work was based on reference tissue input models with reversible kinetics to produce a physically meaningful nonlinearity affecting the TACs of non-specific binding tissues. {Moreover, it considered the mixed kinetics that can be present in each voxel due to partial volume, PSF and biological heterogeneity}. The resulting method managed to recover the binding potential related to the different responses of the tissues to tracer kinetics on simulations. It also provided the tissue affected by abnormalities. The potential interest of this novel technique was evaluated on synthetic and real data, showing promising results. Future works should focus on generalizing this model for settings with irreversible kinetics.

\appendix
\vspace{-0.2cm}

\section{Algo. \ref{algo5:globalplmm} -- Optimization w.r.t. $\mathbf{M}$}
For $k=(1,\cdots,K-1)$, the required gradient is written
\begin{equation}
\setlength{\jot}{0pt}
\begin{split}
\nabla_{m_k}\mathcal{J}(\mathbf{m}_k, \mathbf{A}_k,\mathbf{W}_k,\mathbf{E}_k) = -(\mathbf{\tilde{Y}}-\mathbf{m}_k\mathbf{A}_k)\mathbf{A}_k^T \\ 
-\sum_{i=0}^V\mathbf{E}_{ki}^T\mathbf{\tilde{Y}}\mathbf{W}_{k,i}^T+\sum_{i=0}^V(\mathbf{E}_{ki}+\mathbf{E}_{ki}^T)\mathbf{m}_k\mathbf{W}_{k,i}\mathbf{A}_k^T
\\ 
+\frac{1}{2}\sum_{i=0}^V\sum_{j=0}^V(\mathbf{E}_{ki}^T\mathbf{E}_{kj}+(\mathbf{E}_{ki}^T\mathbf{E}_{kj})^T)(\mathbf{m}_k\mathbf{W}_{k,j}\mathbf{W}_{k,i}^T) \\ 
+\beta(\mathbf{\tilde{M}}-\mathbf{\tilde{M}}^0)
\end{split}
\end{equation}\vspace{-0.2cm}
with $\mathbf{\tilde{Y}}=\mathbf{Y}-\sum_{j \neq k}{\bigg(\mathbf{m}_j\mathbf{A}_j-\sum_{i=0}^V\mathbf{E}_{ji}\mathbf{m}_j\mathbf{W}_{ji}\bigg)}$ and $\mathbf{W}_i = (\mathbf{\tilde{A}}\circ \mathbf{B}_i)$.

The Lipschitz constant is defined as
\begin{equation}
\begin{split}
L_{m_k} =\|\mathbf{A}_k\mathbf{A}_k^T\|+\sum_{i=0}^V\|\mathbf{E}_{ki}+\mathbf{E}_{ki}^T\|\|\mathbf{W}_{k,i}\mathbf{A}_k^T\|\vspace{-0.3cm}
\\+\sum_{i=0}^V\sum_{j=0}^V\|\mathbf{E}_{ki}^T\mathbf{E}_{kj}\|\|\mathbf{W}_{k,j}\mathbf{W}_{k,i}^T\|+\beta\vspace{-0.2cm}
\end{split}
\end{equation}
where the spectral norm $\big\|\mathbf{X}\big\|=\sigma_{\max}(\mathbf{X})$ is the largest singular value of $\mathbf{X}$ and $\big\|\mathbf{X}\big\|_{\infty}=\max_{1\leq i \leq m}\sum_{j=1}^{n}{|x_{ij}|}$ is the sum of the absolute values of the matrix row entries. 
    
For $k=K$, the gradient writes
\begin{equation*}
\nabla_{m_K}\mathcal{J}(\mathbf{m}_K, \mathbf{A}_K) = -(\mathbf{\tilde{Y}}-\mathbf{m}_K\mathbf{A}_K)\mathbf{A}_K^T+\beta(\mathbf{m}_K-\mathbf{m}_K^0)
\end{equation*}
with $\mathbf{\tilde{Y}}=\mathbf{Y}-\mathbf{\tilde{M}}\mathbf{\tilde{A}}-\sum_{i=0}^V \mathbf{Q}_i(\mathbf{\tilde{A}}\circ \mathbf{B}_i)$. The Lipschitz constant is\vspace{-0.2cm}
\begin{equation*}
\begin{split}
L_{m_K} = \|\mathbf{A}_K\mathbf{A}_K^T\|+\beta.\vspace{-0.2cm}
\end{split}
\end{equation*}

\section{Algo. \ref{algo5:globalplmm} -- Optimization w.r.t. $\mathbf{A}$}
For $\mathbf{\tilde{A}}$, the gradient can be computed as
\begin{equation}
\begin{split}
\nabla_{\tilde{A}}\mathcal{J}(\mathbf{\tilde{M}}, \mathbf{\tilde{A}}, \mathbf{Q},\mathbf{B}) = -\mathbf{\tilde{M}}^T\mathbf{\tilde{Y}}
 -\sum_{j=0}^V\bigg((\mathbf{Q}_j^T\mathbf{\tilde{Y}}) \circ \mathbf{B}_j\bigg) \vspace{-0.3cm}
 \\+\eta\mathbf{\tilde{A}}\mathbf{S}\mathbf{S}^T\vspace{-0.3cm}
\end{split}
\end{equation}
with $\mathbf{\tilde{Y}}=\mathbf{Y} - \mathbf{M}\mathbf{A}-\sum_{i=0}^V\mathbf{Q}_i(\mathbf{\tilde{A}}\circ \mathbf{B}_i)$ and $\mathbf{\tilde{M}} = [\mathbf{m}_1,\cdots,\mathbf{m}_{K-1}]$.

The corresponding Lipschitz constant is defined as 
\begin{equation}
\begin{split}
L_{\tilde{A}}= \sum_{i=0}^V\bigg( 2\|\mathbf{\tilde{M}}^T\mathbf{Q}_i\|\|\mathbf{B}_i\|+\|\mathbf{B}_j\|\sum_{j=0}^V\|\mathbf{Q}_j^T\mathbf{Q}_i\|\|\mathbf{B}_i\|\bigg)\vspace{-0.3cm}
\\+\|\mathbf{\tilde{M}}^T\mathbf{\tilde{M}}\|+\eta\big\| \mathbf{S}\mathbf{S}^T\big\|.\vspace{-0.3cm}
\end{split}
\end{equation}
For $\mathbf{A}_K$, the gradient writes
\begin{equation*}
\nabla_{A_K}\mathcal{J}(\mathbf{m}_K, \mathbf{A}_K)  = -\mathbf{m}_K^T\mathbf{\tilde{Y}}+\eta\mathbf{A}_K\mathbf{S}\mathbf{S}^T 
\end{equation*}
with $\mathbf{\tilde{Y}}=\mathbf{Y}-\mathbf{M}\mathbf{A}-\sum_{i=0}^V \mathbf{Q}_i(\mathbf{\tilde{A}}\circ \mathbf{B}_i)$. The Lipschitz constant is
\begin{equation*}
\begin{split}
L_{A_K} =\|\mathbf{m}_K^T\mathbf{m}_K\|+\eta\big\| \mathbf{S}\mathbf{S}^T\big\|.
\end{split}
\end{equation*}

The Lipschitz constant is given by 
\begin{equation}
L_{B_i} = \|\mathbf{Q}_i^T\mathbf{Q}_i\|\|\mathbf{\tilde{A}}\|^2.\vspace{-0.3cm}
\end{equation}

\section{Algo. \ref{algo5:globalplmm} -- Optimization with respect to $\mathbf{B}_i$}
The gradient writes
\vspace{-0.2cm}\begin{equation}
\nabla_{B_i}\mathcal{J}(\mathbf{\tilde{A}}, \mathbf{B},\mathbf{Q}) = -\bigg((\mathbf{Q}_i^T(\mathbf{\tilde{Y}}-\mathbf{Q}_i(\mathbf{\tilde{A}}\circ \mathbf{B}_i)))\circ\mathbf{\tilde{A}}\bigg)
\end{equation}
with $\mathbf{\tilde{Y}} = \mathbf{Y}-\mathbf{M}\mathbf{A}-\sum_{j\neq i}\mathbf{Q}_j(\mathbf{\tilde{A}}\circ \mathbf{B}_j)$.

The Lipschitz constant is
\vspace{-0.2cm}\begin{equation}
L_{B_i} = \|\mathbf{Q}_i^T\mathbf{Q}_i\|\|\mathbf{\tilde{A}}\|^2. \vspace{-0.3cm}
\end{equation}

\section{Algo. \ref{algo5:globalplmm} -- Optimization with respect to $\alpha_{ki}$}

The gradient writes
\vspace{-0.2cm}\begin{equation}
\begin{split}
\nabla_{\alpha_{ki}} \mathcal{J}(\alpha_{ki})=\mathbf{W}_{k,i}(\mathbf{\tilde{Y}}^T(T_p(\mathbf{t})\circ\mathbf{E}_{ki})\\-\frac{1}{2}\mathbf{W}_{k,i}^T\mathbf{m}_k^T((T_p(\mathbf{t})\circ\mathbf{E}_{ki})^T\mathbf{E}_{ki}+\mathbf{E}_{ki}^T(T_p(\mathbf{t})\circ\mathbf{E}_{ki})))\mathbf{m}_k,\vspace{-0.3cm} 
\end{split}
\end{equation}
with $\mathbf{\tilde{Y}}=\mathbf{Y}-\mathbf{M}\mathbf{A}-\sum_{j\neq i}\mathbf{Q}_j\mathbf{W}_j-\sum_{u\neq k}\mathbf{E}_{ui}\mathbf{m}_u\mathbf{w}_{ui}$.

The Lipschitz constant is
\begin{equation}
\begin{split}
L_{\alpha_{ki}} =\|\mathbf{W}_{k,i}\|\bigg(\|-\mathbf{\tilde{Y}}^T+\frac{1}{2}\mathbf{W}_{k,i}^T\mathbf{m}_k^T\mathbf{E}_{ki}^T\|
\\+\frac{3}{2}\|\mathbf{W}_{k,i}^T\mathbf{m}_k^T\|\|\mathbf{E}_{ki}\|\bigg)\|\mathbf{E}_{ki}\|\|T_p(\mathbf{t})\|^2\|\mathbf{m}_k\|.
\end{split}
\end{equation}

\section*{References}
\bibliographystyle{elsarticle-harv} 
\bibliography{string_all_ref,bibli}
\end{document}